\documentstyle[preprint,aps,floats,epsfig]{revtex}
\tightenlines
\begin{document}

\title{$e^+e^-$ production from $pp$ reactions at BEVALAC energies}
\author{E. L. Bratkovskaya, W. Cassing, M. Effenberger
and U. Mosel  \\[2mm]}
\address{Institute f\"ur Theoretiche Physik, Universit\"at
Giessen, 35392 Giessen, Germany}
\maketitle

\begin{abstract}
We have performed a detailed study of dilepton production from $pp$
collisions including the subthreshold $\rho$ production via baryonic
resonances ($N(1520), N(1700)$) in addition to the conventional
dilepton sources as $\pi^0$, $\eta$, $\omega$ and $\Delta$ Dalitz decays
and direct decays of vector mesons ($\rho$, $\omega$).  The role of
baryonic resonances in $\rho$ production from
nucleon-nucleon collisions is studied in comparison to the DLS data
which are well described.
\end{abstract}

\vspace{0.5cm}
PACS: \ {25.75.Dw, 13.30.Ce, 12.40.Yx}

\vspace{0.5cm}
Keywords: particle and resonance production; leptonic and semileptonic
decays; hadron mass models and calculations

\section{Introduction}
Dileptons are the most clear probes for an investigation of the early
phases of high-energy heavy-ion collisions because they may leave the
reaction volume essentially undistorted by final-state interactions.
Dilepton spectra from heavy-ion collisions can provide information
about the restoration of chiral symmetry and in-medium properties of
hadrons (cf. Refs. \cite{BrownRho,H&L92,Shakin94,Klingl96,Asakawa93,mosel91}).
According to QCD sum rules \cite{H&L92,Asakawa93,Leupold} as
well as QCD inspired effective Lagrangian models
\cite{BrownRho,Shakin94,Klingl96,Herrmann},\cite{asakawa,Chanfray,Rapp,Friman,RappNPA,Peters}
the vector mesons ($\rho$, $\omega$ and $\phi$) significantly change
their properties with the nuclear density.
In fact, the experimentally observed enhanced production of soft lepton
pairs above known sources in nucleus-nucleus collisions at SPS
energies \cite{CERES,HELIOS} might be due to the in-medium
modification of vector mesons \cite{Li,Li96,Cass95CH,Brat97,CBRW97}.

Dileptons from heavy-ion collisions have also been measured
by the DLS Collaboration \cite{DLSold,DLSnew} at BEVALAC energies,
where a different temperature and density regime is probed.
The first generation of DLS data \cite{DLSold}, based on a limited
data set, were consistent with the results from transport model
calculations \cite{Xiong90,Wolf90,Gudima,BCMas96} including the
conventional dileptons sources as $pn$ bremsstrahlung, $\pi^0$, $\eta$,
$\omega$ and $\Delta$ Dalitz decay, direct decay of vector mesons
and pion-pion annihilation, without incorporating any medium effects.
A recent DLS measurement \cite{DLSnew} including the full data set and
an improved analysis shows an increase by about a factor of 5-7 in the
cross section in comparison to Ref. \cite{DLSold}
and the early theoretical predictions.

In Ref. \cite{BCRW97} the in-medium rho spectral functions from
Refs.~\cite{RappNPA,Peters} have been implemented in the HSD transport
approach for $\rho$ mesons produced in baryon-baryon, pion-baryon
collisions as well as from $\pi\pi$ annihilation, and a
factor of 2-3 enhancement has been obtained compared to the case of
a free $\rho$-spectral function. In Ref.~\cite{Ernst}
dropping vector meson masses and $\omega$ meson broadening were
incorporated in the UrQMD transport model, which also showed an enhancement
of the dilepton yield, however, the authors could not describe the new DLS data
\cite{DLSnew} as well.
Another attempt to solve the DLS 'puzzle' has been performed recently in
Ref.~\cite{BrKo98} where the dropping hadron mass scenario was
considered together with the subthreshold $\rho$ production in
$\pi N$ scattering via the baryonic resonance $N(1520)$ whose
importance was pointed out in Refs.~\cite{Peters,Stony98}.
It was found that the enhancement of the dilepton spectra due to low
mass $\rho$'s from the $N(1520)$ was not sufficient to match the DLS data.
Thus, all in-medium scenarios that successfully have explained the dilepton
enhancement at SPS energies  failed to describe the new DLS
dilepton data \cite{DLSnew} from heavy-ion collisions.

Recently the DLS Collaboration has published dilepton data from
elementary $pp$ and $pd$ collisions at 1-5~GeV \cite{DLSpp}.  This
provides the possibility for an independent check of the elementary
dilepton channels that enter as 'input' for transport calculations for
heavy-ion reactions.  Such an analysis has been carried out in
Ref.~\cite{Ernst} and it was shown that the dilepton invariant mass
spectra from $pp$ reactions at invariant masses from $0.3 \le M\le
0.6$~GeV are underestimated when incorporating only the $\pi^0$,
$\eta$, $\omega$ and $\Delta$ Dalitz decays and direct decays of vector
mesons ($\rho$, $\omega$).

Historically, the first calculations of dilepton production from
hadronic bremsstrahlung and from the radiative decay of the $\Delta$
resonance in nucleon-nucleon collisions have been based on the
soft-photon approximation \cite{Gale87} which can be considered
as an upper limit \cite{Lichard}.  Later on, the model has been
refined and the dilepton production from elementary nucleon-nucleon
reactions was studied within different microscopic models
\cite{Schaef89,Schaefer,Haglin91,BrTit,Fred}.  In
Ref.~\cite{Schaef89,Schaefer} the dilepton yield was calculated at 1.0
and 2.1 GeV on the basis of the one-boson exchange model taking into
account the dilepton production via $\Delta$ Dalitz decay and
bremsstrahlung.  It was found that the incorporation of the
vector-meson form factor into the vertices with a $\Delta, N$ or $\pi$
leads to a significant enhancement of the dilepton yield in the $\rho$
mass region which substantially overestimated the $p Be$ data
\cite{DLSold} at that time. The same effect was found for $pp$ dilepton
radiation at 1 GeV in Ref.  \cite{Fred} based on a T-matrix approach.
Furthermore, the dilepton yield from $pp$ collisions at 4.9 GeV was
calculated in Ref.~\cite{Haglin93} and it was found that 'inelastic'
channels (final states with one or two pions in addition to the
nucleons and dileptons) dominated the bremsstrahlung calculated within
the soft-photon approximation.

In this article we perform a detailed study of dilepton production from
$pp$ collisions including the subthreshold $\rho$ production via
baryonic resonances ($N(1520), N(1700)$) in addition to the
conventional dilepton sources ($\pi^0$, $\eta$, $\omega$ and $\Delta$
Dalitz decay, direct decay of vector mesons ($\rho$, $\omega$)).  We
investigate the role of baryonic resonances in $\rho$ production from
nucleon-nucleon collisions on the basis of two independent resonance
models from Peters et al. \cite{Peters} and Manley et al.
\cite{Manley} in order to demonstrate the dominant theoretical uncertainties.

The paper is organized as follows:  In Section~2 we discuss the $\rho$
meson production via baryonic resonances in nucleon-nucleon collisions.
In Section~3 we describe the elementary dilepton production channels
while Section 4 contains a comparison with the DLS data. We close
with a summary and discussion of open problems in Section 5.

\section{$\rho$ meson production via baryon resonances in
nucleon-nucleon collisions}

In the resonance model the total $\rho$ production cross section from
$N N$ collisions via the N(1520) can be calculated as
\begin{eqnarray}
\sigma(s)^{NN\to RN \to \rho NN} = \int\limits_{m_N+m_\pi}^{\sqrt{s}-m_N}
d\mu \ \sigma(s,\mu)^{N N \to RN} \times {2\over \pi}
{\mu^2\over (\mu^2-m_R^2)^2+(\mu \Gamma_{tot}^R(\mu))^2} \
{\Gamma(\mu)^{R\to\rho N}}.
\label{NNrhos}\end{eqnarray}
The resonance production cross section $\sigma(s,\mu)^{NN \to RN}$
with good accuracy can
be approximated by a constant matrix element (cf. Ref.~\cite{TeisZP97})
and a flux factor, i.e.
\begin{eqnarray}
&&\sigma(s,\mu)^{N N \to RN} = {p_f\over s p_i}
{|\bar{\cal M}|^2\over  16\pi}, \label{nnrho}\\
&& p_f = {\left[(s -(\mu + m_N)^2) (s -(\mu -m_N)^2)\right]^{1/2}
\over 2 \sqrt{s}}, \  \
p_i = {(s-4m_N^2)^{1/2} \over 2}.
\label{momres}\end{eqnarray}
The average matrix element $|\bar{\cal M}|$
for the $pp\to N^+(1520)p$ reaction
has been obtained in Ref.~\cite{TeisZP97} from a fit to $\pi$- and
$\rho$- production channels; it was found to be ${|\bar{\cal M}|}^2 =
64 \pi$~mb$\cdot$GeV$^2$.

According to Ref.~\cite{Peters} the partial decay width for the resonance
decay ${R\to \pi N}$ is taken as
\begin{eqnarray}
&&\Gamma_{R\to \pi N} (\mu) = \Gamma_0 \left(k_\pi(\mu)\over k_\pi(m_R)\right)^{2l+1} \
\times \left(0.5^2+{k_\pi(m_R)}^2 \over 0.5^2+{k_\pi(\mu)}^2\right)^{2l+1},
	\label{partW} \\
&& k_\pi(\mu) = {\left[(\mu^2 -(m_\pi + m_N)^2)
(\mu^2 -(m_\pi -m_N)^2)\right]^{1/2}  \over 2 \mu},
\end{eqnarray}
with $l=2$ and $\Gamma_0=0.066$~GeV for the $N(1520)$ resonance.

The total resonance decay width is assumed to be given by the sum of the
pion- and rho-decay width (cf. page 116 of Ref. \cite{Peters}):
\begin{eqnarray}
\Gamma_{tot}^R =  \Gamma_{R\to \pi N} + \Gamma_{R\to \rho N},
\label{GRtot}\end{eqnarray}
where $\Gamma_{R\to \pi N}$ is defined according to Eq.~(\ref{partW})
with $\Gamma_0=0.095$~GeV for $N(1520)$. The width $\Gamma_{R\to \rho N}$
at an invariant mass $\mu$ is calculated as
\begin{eqnarray}
 \Gamma_{R\to \rho N} (\mu) = \int\limits_{2m_\pi}^{\mu-m_N} dM
 {d\Gamma(\mu,M)\over dM}^{R\to \rho N}.
\label{IntGrho}\end{eqnarray}
Furthermore, the partial decay width for the process
${R\to \rho N}$ is defined
according to Ref.\cite{Peters} in the following way:
\begin{eqnarray}
&& {d\Gamma(\mu,M)\over dM}^{R\to \rho N} =
\left({f_{RN\rho}\over m_\rho}\right)^2 \ {1\over \pi} \
M {m_N\over \mu} \  k_\rho \ (2\omega_\rho^2+M^2) \
\ A(M)\ F(k_\rho^2),
\label{GKo}\end{eqnarray}
where $\mu$ is the resonance mass, $f_{RN\rho}$ is the coupling
constant ($f_{N(1520)N\rho}=7$ \cite{Peters}), $k_\rho$ denotes
the three-momentum of the $\rho$ meson in the $R$ resonance rest frame,
\begin{eqnarray}
k_\rho = {\left[(\mu -(M + m_N)^2) (\mu -(M -m_N)^2)\right]^{1/2}
\over 2 \mu}, \  \
\label{Krho}\end{eqnarray}
while $\omega_\rho^2=k_\rho^2 + M^2$ is the $\rho$ energy. In Eq.~(\ref{GKo})
$A(M)$ is the vacuum $\rho$ spectral function, i.e. a Breit-Wigner
distribution
\begin{eqnarray}
A(M) = {1\over \pi} {m_\rho \Gamma_{tot}^\rho(M) \over
(M^2-m_\rho^2)^2 + (m_\rho {\Gamma_{tot}^{\rho}(M)})^2}.
\label{BW}\end{eqnarray}
Moreover, $F({\bf k}_\rho^2)$ is a monopol form factor
\begin{eqnarray}
 F({\bf k}_\rho^2) ={\Lambda^2\over \Lambda^2+{\bf k_\rho}^2},
\ \Lambda=1.5 \ {\rm GeV},
\label{fff}\end{eqnarray}
while the total $\rho$ width is
\begin{eqnarray}
\Gamma_{tot}^\rho(M) = \Gamma_{\rho\to \pi\pi}(m_\rho) \
 \left(m_\rho\over M\right) \ \left(k_\pi(M)\over k_\pi(m_\rho)\right)^3.
\label{rhowidth}\end{eqnarray}

\begin{figure}
\epsfig{file=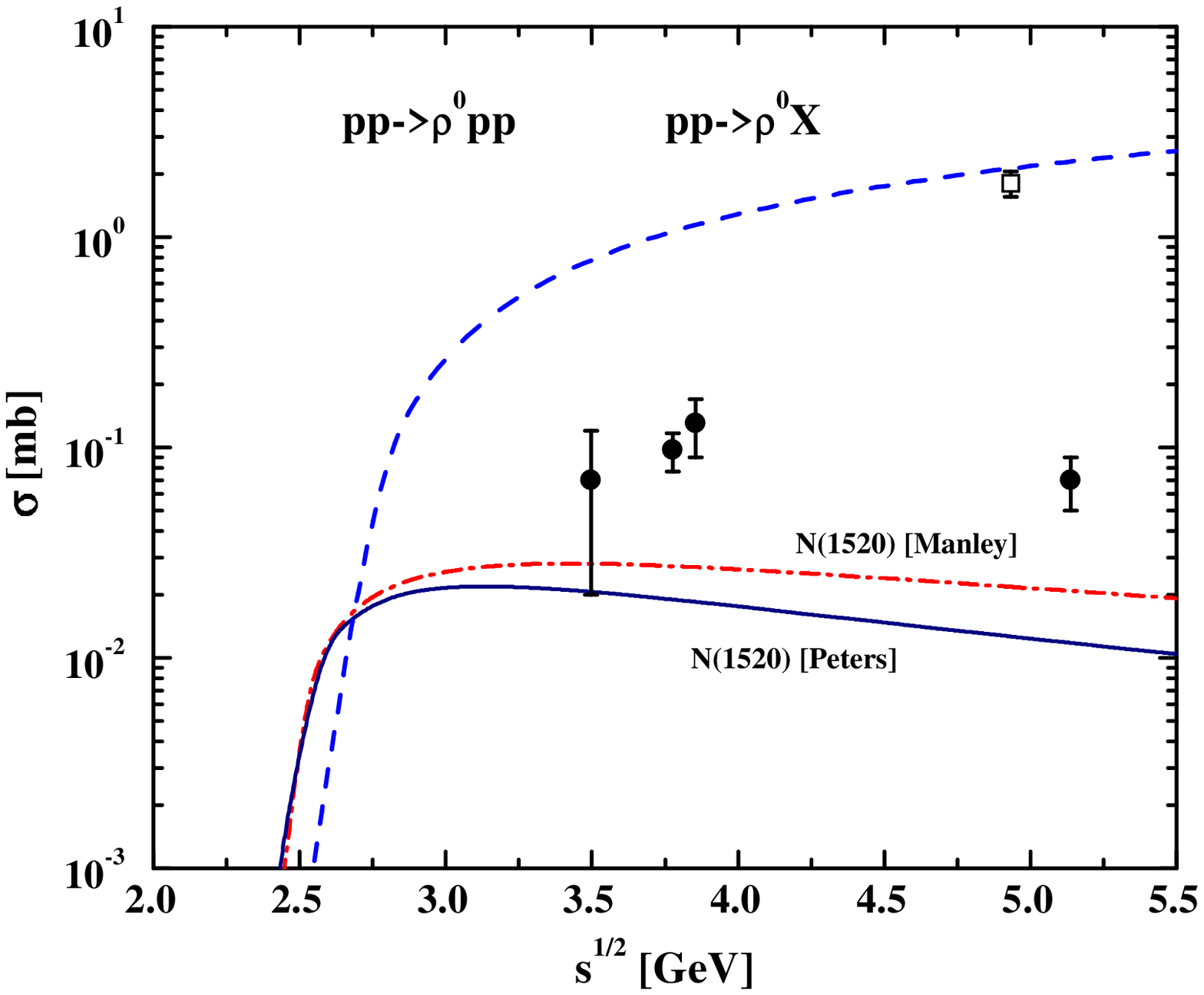,width=15cm}
\vspace*{-2cm}
\caption{The $\rho$ production cross section from the $pp \to N(1520)p \to
\rho^0 pp$ channel calculated according to the resonance models from
Ref.~\protect\cite{Peters} (full line) and from Ref.~\protect\cite{Manley}
(dot-dashed line) as a function of the invariant energy $\sqrt{s}$.
The dashed line indicates the parametrization for inclusive $\rho$
production  in $pp$ collisions from Ref.~\protect\cite{SibCM97}
integrated over the kinematically allowed range of invariant mass 
of the $\rho$ spectral function.
The full circles are the experimental data \protect\cite{LB} for the
exclusive $\rho$ production whereas the open square corresponds to the
inclusive data point at high $\sqrt{s}$.}
\label{Fig1pp}
\end{figure}

The $\rho$ production cross section from $pp$ collisions via $N(1520)$
calculated according to Eqs.~(\ref{NNrhos})--(\ref{momres}) with the
resonance widths from Peters et al. \cite{Peters} is shown in
Fig.~\ref{Fig1pp} as a solid line. The dot-dashed line presents the
result within the analysis from Manley et al. \cite{Manley}.  The dashed
line, furthermore, indicates the parametrization for the inclusive
$\rho$ production  from Ref.~\cite{SibCM97} integrated over the
kinematically allowed range of invariant mass of the $\rho$
spectral function.
The full circles are the
experimental data \cite{LB} for the exclusive $\rho$ production whereas
the open square corresponds to the inclusive data point at high
$\sqrt{s}$.

As seen from Fig.~\ref{Fig1pp} at $\sqrt{s} > 2 m_N + m_\rho$ the
relative contribution of $\rho$'s stemming from the $N(1520)$ is small
compared to the total inclusive $\rho$ production cross section.
However, closer to threshold the $N(1520)$
plays a dominant role in $\rho$ production with low invariant masses
since the cross section is dominted by the resonance amplitude.

In our analysis we also include the $N(1700)$ resonance in a similar
way as the $N(1520)$ using again the parameters from Peters et al.
\cite{Peters}. We discard higher resonances, e.g. $N(1720), N(1905)$,
because already the $N(1700)$ plays only a minor role in the results
to be discussed in this study.


\section{Elementary dilepton production channels}

\begin{figure}
\epsfig{file=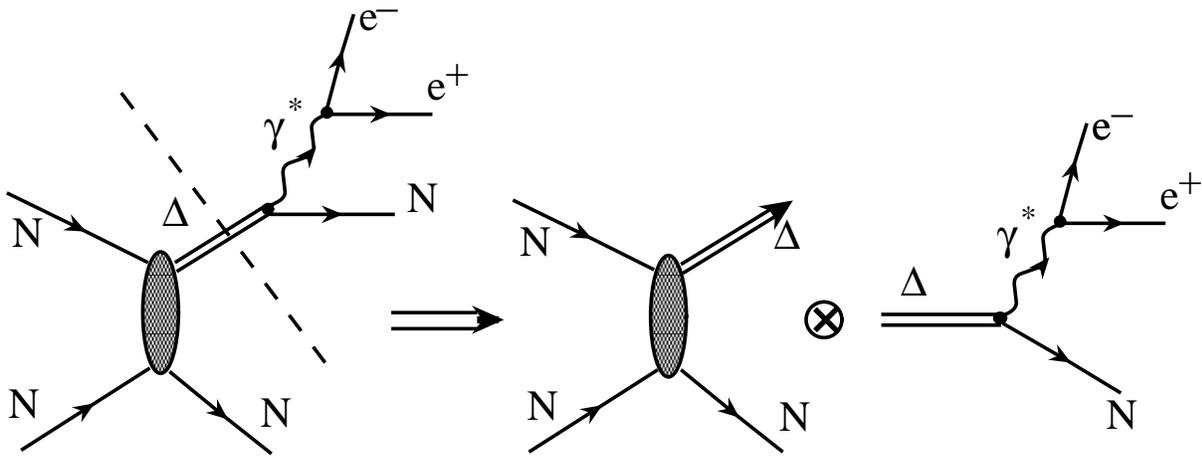,width=16cm}
\caption{Factorization of the diagram for the process
(\ref{ch3}) with an intermediate $\Delta$.}
\label{Fig2pp}
\end{figure}

In our analysis we calculate dilepton production by taking into account
the contributions from the following channels:
\begin{eqnarray}
&& pp \to N(1520) p, \ N(1520) \to \rho^0 p \to e^+e^- p, \label{ch1} \\
&& pp \to N(1700) p, \ N(1700) \to \rho^0 p \to e^+e^- p, \label{ch2} \\
&& pp \to \Delta N, \ \Delta \to N e^+e^-,       \label{ch3} \\
&& pp \to \eta X,   \ \eta \to \gamma e^+e^-,    \label{ch4} \\
&& pp \to \omega X, \ \omega \to \pi^0 e^+e^-,   \label{ch5} \\
&& pp \to \pi^0 X, \  \pi^0 \to \gamma e^+e^-,  \label{chpi0} \\
&& pp \to \rho^0 X, \ \rho \to e^+e^-,           \label{ch6} \\
&& pp \to \omega X, \ \omega \to e^+e^-.         \label{ch7}
\end{eqnarray}
For the processes (\ref{ch1})--(\ref{ch7}) we use the assumption that
the amplitude can be factorized into a meson or resonance production and
dilepton decay part, i.e. we cut the diagram as illustrated in
Fig.~\ref{Fig2pp}, e.g., for the process (\ref{ch3}) with an intermediate
$\Delta$ resonance. This assumption provides not only a significant
simplification, which is necessary for applications in transport
calculations \cite{Wolf90,CB99PR}, but also allows to take into account
the 'inelasticity' from many-particle production channels which become
dominant at high energy. Note that the factorization ansatz holds (also in
case of broad resonances) when integrating over the final relative angle
between the dilepton and a proton; it becomes questionable only for the
multi-differential dilepton cross sections. Since we will compare to
practically angle integrated data (see below), this approximation should
be justified for our present study.

We discard the $pp$ bremsstrahlung from the nucleon pole since
the microscopic OBE calculations \cite{Schaefer,BrTit,Fred}
have shown that at 1.0 GeV the $pp$ bremsstrahlung is smaller than
the $\Delta$ Dalitz decay contribution by a factor of 2-3.
At this energy also the interference between nucleon and $\Delta$-pole
terms is negligible. At high energies, where this interference
becomes important \cite{Schaefer}, the overall contribution from
these channels is negligible (see below).

\subsection{Dilepton production through the $N(1520)$ resonance}

Within the above assumptions the dilepton cross section from $NN$
collisions via the $N(1520)$ can be factorized as
\begin{eqnarray}
{d\sigma(s,M)\over dM}^{N N\to RN \to \rho^0 NN \to e^+e^-NN}  =
{d\sigma(s,M)\over dM}^{N N\to \rho NN} \
{\Gamma_{\rho\to e^+e^-}(M) \over \Gamma_{tot}^\rho(M)},
\label{NNdil} \end{eqnarray}
where the differential $\rho$ production cross section with
mass $M$ is
\begin{eqnarray}
{d\sigma(s,M)\over dM}^{N N\to \rho NN} &&=
\int\limits_{m_N+m_\pi}^{\sqrt{s}-m_N} d\mu \
\sigma(s,\mu)^{N N \to RN} \nonumber \\
&&\times {2\over \pi}
{\mu^2\over (\mu^2-m_R^2)^2+(\mu \Gamma_{tot}^R(\mu))^2} \
{ d\Gamma(\mu,M)^{R\to\rho N}\over dM}.
\label{NNrho}\end{eqnarray}
The partial decay width ${ d\Gamma(\mu,M)^{R\to\rho N} / dM}$ is
given by Eq.~(\ref{GKo}), the total width $\Gamma^R_{tot}$
is defined according Eq. (\ref{GRtot}) while the $\rho$ width
is taken according to Eq.~(\ref{rhowidth}).

In (\ref{NNdil}), $\Gamma_{\rho\to e^+e^-}(M)$ is the dilepton decay width
of a neutral $\rho$ meson of mass $M$ which in line with the vector
dominance model is taken as
\begin{equation}
\Gamma_{\rho \to e^+e^-}(M)= C_\rho \ \frac{m_\rho^4}{M^3}
\label{ree}\end{equation}
with $C_\rho=8.8\times 10^{-6}$.

To compare with the experimental data of the DLS collaboration one has
to use the appropriate experimental filter, which is a function of the
dilepton invariant mass $M$, transverse momentum $q_T$ and rapidity
$y_{lab}$ in the laboratory frame  -- $F(M,q_T,y_{lab})$. For that
purpose we explicitly simulate the dileptons by Monte-Carlo: first,
we produce the resonance $N(1520)$ isotropically in the center-of-mass
frame of the $pp$ collision with momentum $p_f$ (\ref{momres}). In the
same way we create the $\rho$ mesons in the $N(1520)$ rest frame with
momentum $k_\rho$ (\ref{Krho}) and perform a Lorentz transformation to
the laboratory frame, where the filter $F(M,q_T,y_{lab})$ is applied.  A
similar procedure has been used for all other channels.
We note that we applied the DLS acceptance filter (version 4.1).

\begin{figure}
\epsfig{file=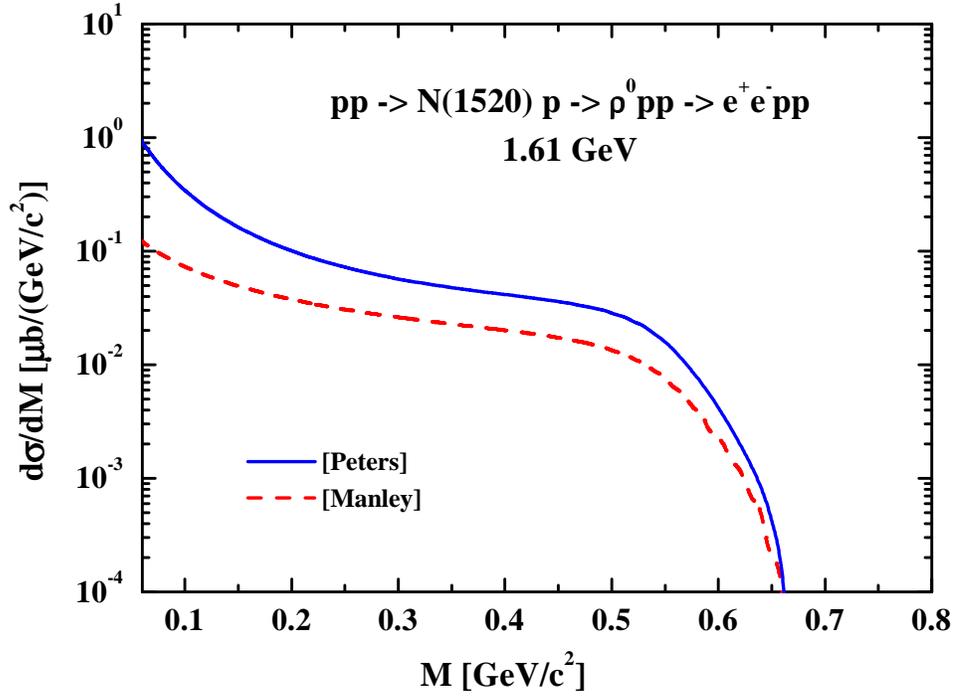,width=15cm}
\caption{The dilepton invariant mass spectra from the channel
$pp\to N(1520)p \to \rho^0 pp$ $\to e^+e^- pp$ at 1.61~GeV calculated
according to the resonance model of Peters et al. \protect\cite{Peters}
(solid line) and Manley et al. \protect\cite{Manley} (dashed line)
without implementing an experimental filter.}
\label{Fig3pp}
\end{figure}

In Fig.~\ref{Fig3pp} we show the dilepton invariant mass spectra from
the channel $pp \to N(1520) p \to \rho^0 pp \to e^+e^- pp$ at 1.61~GeV
calculated without experimental filter. The solid line indicates the
result calculated according to Peters et al. \cite{Peters}, while the
dashed line corresponds to the resonance analysis of Manley et al.
\cite{Manley}. The dilepton yield for the parameters from \cite{Peters}
is by a factor of 2 larger then that from  Manley et al. \cite{Manley}
due to the differences in the resonance parameters and parametrizations
for the widths. Our future results for dileptons via $N(1520)$ are
based on Ref.~\cite{Peters}, however, one has to keep in mind the
theoretical uncertainties indicated here. This also holds for the
uncertainties induced by the neglect of interference terms between the
different diagrams. Whereas the latter vanish for diagrams with different
quantum numbers in the intermediate state after integration over the
relative angle between the dilepton and a final proton, they might be
essential in some regions of phase space for multi-differential cross
sections. However, when integrating over a wide region of phase space
(see below) the uncertainty to due the interference of diagrams should be
much smaller than the uncertainty of the factor of 2 stemming
from the different resonance model parameters.

\subsection{Dalitz decays}

The process $pp \to \Delta N \to NN e^+e^-$ is treated as a two step
process -- the $\Delta$ production from the $pp$ interaction ($pp \to
\Delta N$) and the $\Delta$ Dalitz decay ($\Delta \to N e^+e^-$) -- cf.
Fig.~\ref{Fig2pp}. For the $\Delta$ production below 2.1 GeV we adopt
the differential cross section from Refs.~\cite{Wolf90,TeisZP97}, where
the $\Delta$ resonances are created with a mass according to a
Breit-Wigner distribution with a momentum dependent width.
At the higher energies we obtain the $\Delta$ cross section from the
LUND model \cite{FRITIOF}.
Also the $\pi^0$ production has been performed via the
$\Delta$ resonance excitation and decay in line with
Refs.~\cite{Wolf90,TeisZP97}.

For the $\Delta$ Dalitz-decay we use the $N \Delta \gamma$ vertex
as in Ref.~\cite{Wolf90}
\begin{equation}
{\cal L}_{int} = e A^\mu {\bar \Psi}^\beta_{\Delta} \Gamma_{\beta \mu} \Psi_N,
\label{vertex1}
\end{equation}
where
\begin{eqnarray}
&& \Gamma_{\beta\mu} = g \ f \ \eta_{\beta\mu}, \label{GamDelta}\\
&& f = -{3\over 2} {m_\Delta + m_N \over m_N
     \left((m_\Delta + m_N)^2 - M^2\right)}, \nonumber\\
&& \eta_{\beta\mu} = -M\chi_{\beta\mu}^1 + \chi_{\beta\mu}^2
	+ 0.5 \chi_{\beta\mu}^3, \nonumber\\
&& \chi_{\beta\mu}^1 = (q_\beta \gamma_\mu - q_\nu \gamma^\nu g_{\beta\mu})
	\gamma_5, \nonumber\\
&& \chi_{\beta\mu}^2 = (q_\beta {\bar P}_\mu - q_\nu {\bar P}^\nu
	g_{\beta\mu})	\gamma_5, \nonumber\\
&& \chi_{\beta\mu}^3 = (q_\beta q_\mu - M^2 g_{\beta\mu}) \gamma_5, \nonumber\\
&& \bar P = {1\over 2} (p_\Delta +p_N), \nonumber
\end{eqnarray}
and $g = 5.44$ is the coupling constant fitted to the radiative decay
width $\Gamma_0 (0) = 0.72$~MeV.

The processes (\ref{ch4}),(\ref{ch5}),(\ref{chpi0}) are calculated in a
similar way as the $\Delta$ Dalitz decay (\ref{ch3}), i.e. first
$\eta$, $\omega$ or $\pi^0$ mesons are produced in $pp$ interactions and
then their Dalitz decay $\eta \to \gamma e^+e^-$,
$\omega\to \pi^0 e^+e^-$ and $\pi^0\to \gamma e^+e^-$  is simulated by
Monte-Carlo.

For the $\eta$ meson production cross section  we adopt the
parametrization from Refs.~\cite{Wolf90,Vetter}
in line with the data from the WASA collaboration~\cite{WASA}.
The $\eta$ Dalitz-decay to $\gamma e^+e^-$ is given by  \cite{Landsberg}:
\begin{eqnarray}
{d\Gamma_{\eta \to \gamma e^+e^-}\over dM}
= {4\alpha\over 3\pi} \ {\Gamma_{\eta \to 2\gamma}\over M}
\left(1 - {4 m_e^2\over M^2} \right)^{1/2} \left( 1 + 2 {m_e^2\over M^2}
\right)   \nonumber \\
\times \left( 1 - {M^2\over m_\eta^2}\right)^3
\ |F_{\eta \to \gamma e^+e^-}(M)|^2,
\label{gameta}
\end{eqnarray}
where the form factor is parametrized in the pole approximation as
\begin{eqnarray}
F_{\eta \to \gamma e^+e^-} (M) = \left(1-{M^2\over
\Lambda_\eta^2}\right)^{-1}
\label{feta}\end{eqnarray}
with the cut-off parameter $\Lambda_\eta \simeq 0.72$~GeV.

The $\pi^0$ Dalitz decay is calculated in a similar way using
the form factor from Ref.~\cite{Landsberg}, i.e.
\begin{eqnarray}
F_{\pi^0 \to \gamma e^+e^-} (M) = \left(1+ B_{\pi^0} M^2\right),
\label{pi0}\end{eqnarray}
with $B_{\pi^0} = 5.5$~GeV$^{-2}$.

Similarly, the $\omega$ Dalitz-decay is \cite{Landsberg}
\begin{eqnarray}
&& {d\Gamma_{\omega \to \pi^0 e^+e^-}\over dM}
= {2\alpha\over 3\pi} \ {\Gamma_{\omega \to \pi^0 \gamma}\over M}
\left(1 - {4 m_e^2\over M^2} \right)^{1/2} \left( 1 + 2 {m_e^2\over M^2}
\right)   \nonumber \\
&&\times \left[ \left(1+ {M^2\over m_\omega^2 - m_\pi^2}\right)^2 -
{4 m_\omega^2 M^2\over (m_\omega^2-m_\pi^2)^2} \right]^{3/2} \
|F_{\omega \to \pi^0 e^+e^-}(M)|^2,
\label{gamomeg}
\end{eqnarray}
where the form factor squared is parametrized as \cite{Landsberg,Cass95CH}
\begin{equation}
|F_{\omega \to \pi^0 e^+e^-}(M)|^2 = {\Lambda_\omega^4\over
(\Lambda_\omega^2- M^2)^2 + \Lambda_\omega^2 \Gamma_\omega^2}
\label{fomegD}
\end{equation}
with
$$\Lambda_\omega= 0.65\ {\rm GeV}, \ \ \Gamma_\omega = 75\ {\rm MeV}.$$

\subsection{Vector meson decay}

In a first step we calculate the production of vector mesons in $pp$
collisions and as a second step the direct decay of vector mesons to
dileptons.
The vector meson production in $pp$ interactions is evaluated in the
following way: close to threshold, i.e. $T_{kin}\le 2.1$~GeV, we
use the inclusive mass differential parametrization for the $\rho$
(dashed line in Fig.~\ref{Fig1pp}) and $\omega$ production from $pp$
collisions.
The mass of the vector meson $M$ is distributed according to
the Breit-Wigner form:
\begin{equation}
f(M) =  N_V \ {2\over \pi} \ {M m_V \Gamma^V_{tot}
\over (M^2-m_V^2)^2 + (m_V {\Gamma_{tot}^V})^2},
\label{fmdist}\end{equation}
while $N_V$ guarantees normalization to unity, i.e. $\int f(M) dM =1$.
The total $\rho$ meson width $\Gamma_{tot}^\rho$ is defined according
to Eq.~(\ref{rhowidth}).  For the narrow $\omega$ meson we use a
constant width $\Gamma_{tot}^\omega=0.00841$~GeV.
The $\rho$ momentum is simulated according to the 3-body phase space
since we are close to the $\rho$ production threshold.

The vector meson decay to dileptons then is calculated according to
the branching ratio,
\begin{eqnarray}
{\rm Br}(M) = {\Gamma_{V\to e^+e^-}(M) \over \Gamma_{tot}^V(M)},
\label{brrho}\end{eqnarray}
where $\Gamma_{V\to e^+e^-}(M)$ is given by Eq.~(\ref{ree})
with $C_\omega=1.344\times 10^{-6}$ for $\omega$ mesons.

\subsection{Dilepton production at 4.9 GeV}

With increasing energy, i.e. at 4.9 GeV, multiple particle production
channels become dominant. In order to take into account correctly the
many-body phase space, the FRITIOF event generator version 7.02 based
on the LUND string fragmentation model \cite{FRITIOF} has been used for
$\eta, \omega, \rho, \Delta, \pi^0$ production.  The FRITIOF model
gives a good description of the experimental data for meson production
in $NN$ collisions (for details see, e.g., Ref.~\cite{CB99PR}).
The vector mesons as well as $\Delta$ resonances
produced by the FRITIOF generator acquire masses according to the
Breit-Wigner distribution while taking into account the proper
available phase space.
The dilepton decay of $\eta, \omega, \rho, \Delta, \pi^0$ is then
treated in the same way as described above.

\section{Comparison to the DLS data}

\subsection{Differential mass spectra}

In Fig.~\ref{Fig4pp}  we present the calculated dilepton invariant mass
spectra $d\sigma/dM$ for $pp$ collisions from 1.0 -- 4.9 GeV including
the final mass resolution and filter $F(M,q_T,y_{lab})$ from the DLS
collaboration in comparison to the DLS data \cite{DLSpp}.
The thin lines indicate the individual contributions from the different
production channels; {\it i.e.}~ starting from low $M$:
Dalitz decay $\pi^0 \to \gamma e^+ e^-$ (short dashed line),
$\eta \to \gamma e^+ e^-$ (dotted line),
$\Delta \to N e^+ e^-$ (dashed line),
$\omega \to \pi^0 e^+ e^-$ (dot-dot-dashed line),
$N(1520) \to N e^+ e^-$ (dot-dashed line);
$N(1700) \to N e^+ e^-$ (dotted line);
for $M \approx $ 0.7 GeV: $\omega \to e^+e^-$ (dot-dot-dashed line),
$\rho^0 \to e^+e^-$ (short dashed line).
The full solid line represents the sum of all sources considered here.

Whereas at 1.04 GeV the dileptons stem practically all from $\pi^0$ and
$\Delta$ Dalitz decays, $\eta$ and $N(1520)$ Dalitz decays become more
important at 1.27 GeV, which is just above the $\eta$ production
threshold. The contribution from the $N(1520)$ is most prominent at
1.61 GeV and much larger than the other channels for $M \geq$ 0.5 GeV,
whereas the $\eta$ decay dominates already at 1.85 GeV. Note that the
$\eta$ cross section in $pp$ collisions is well known experimentally
\cite{Pinot} as well as the $\pi^0$ and $\Delta$ resonance yields. The
contribution of the $N(1520)$ thus is necessary for a proper
description of the $pp$ data especially at 1.61 GeV. The contribution
of the $N(1700)$ is always below the $N(1520)$ and is practically
not seen.

At 2.1 GeV  the direct
decays of $\rho$ and $\omega$ mesons become dominant for $M\ge 0.65$
GeV.  However, even at this higher energy for $M\simeq 0.6$ GeV the
contribution from the $N(1520)$ is still seen.  At 4.9 GeV the dilepton
yield is dominated by the $\eta$ Dalitz decay and direct
decays of $\rho$ and $\omega$ mesons.  The $\rho$ spectrum is
enhanced towards low $M$ due to the limited phase space and strong
mass dependence $(M^{-3})$ of the dilepton decay width (\ref{ree}) in
line with the vector dominance model.

\subsection{Rapidity spectra}

\begin{figure}
\phantom{a}\vspace*{-1cm}
\epsfig{file=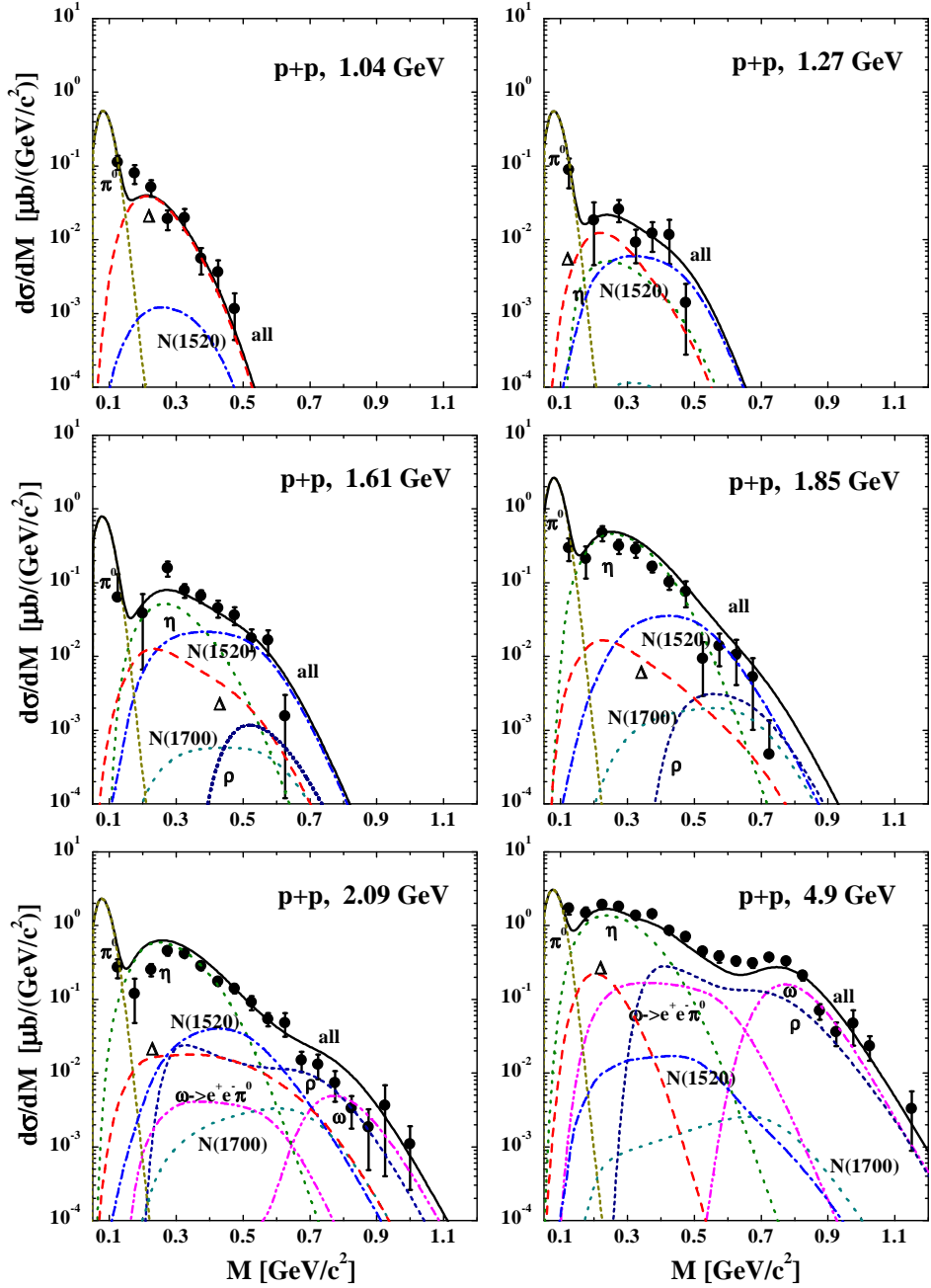,width=15cm}
\vspace*{-1.5cm}
\caption{ The calculated dilepton invariant mass spectra $d\sigma/dM$
for $pp$ collisions from 1.0 -- 4.9 GeV in comparison to the DLS data
\protect\cite{DLSpp}.  The thin lines indicate the individual
contributions from the different production channels; {\it i.e.}~
starting from low $M$:  Dalitz decay $\pi^0 \to \gamma e^+ e^-$ (short
dashed line), $\eta \to \gamma e^+ e^-$ (dotted line), $\Delta \to N
e^+ e^-$ (dashed line), $\omega \to \pi^0 e^+ e^-$ (dot-dot-dashed
line), $N(1520) \to N e^+ e^-$ (dot-dashed line); $N(1700) \to N e^+
e^-$ (dotted line); for $M \approx $ 0.7 GeV: $\omega \to e^+e^-$
(dot-dot-dashed line), $\rho^0 \to e^+e^-$ (short dashed line).  The
full solid line represents the sum of all sources considered here. }
\label{Fig4pp}
\end{figure}

\begin{figure}
\epsfig{file=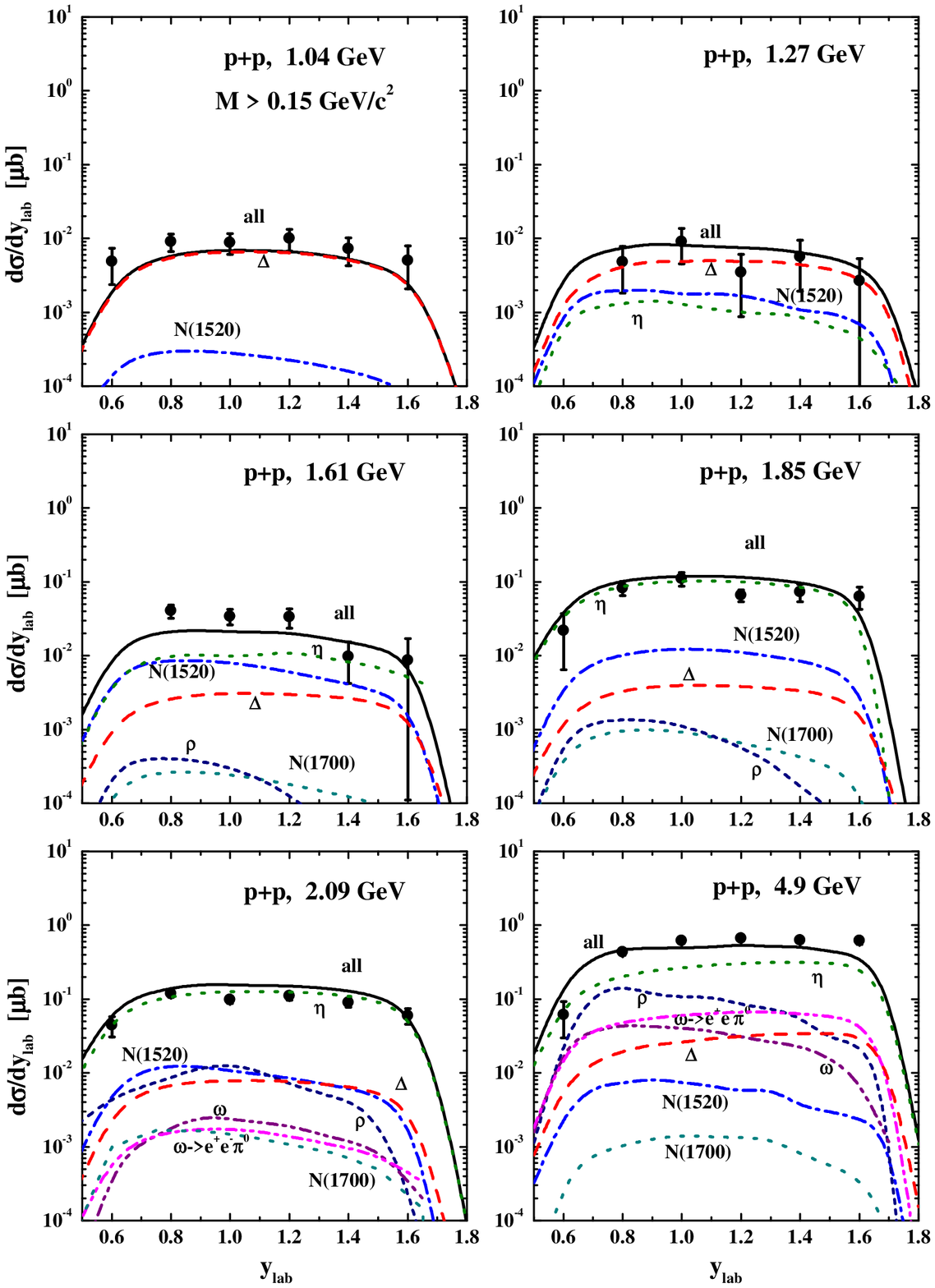,width=15cm}
\caption{The laboratory rapidity spectra of dileptons for $pp$ collisions
at 1.0 -- 4.9~GeV in comparison to the DLS data~\protect\cite{DLSpp}
with a low mass cut of 0.15 GeV/$c^2$.
The assignment of the lines is the same as in Fig. \protect\ref{Fig4pp}.}
\label{Fig5pp}
\end{figure}

\begin{figure}
\epsfig{file=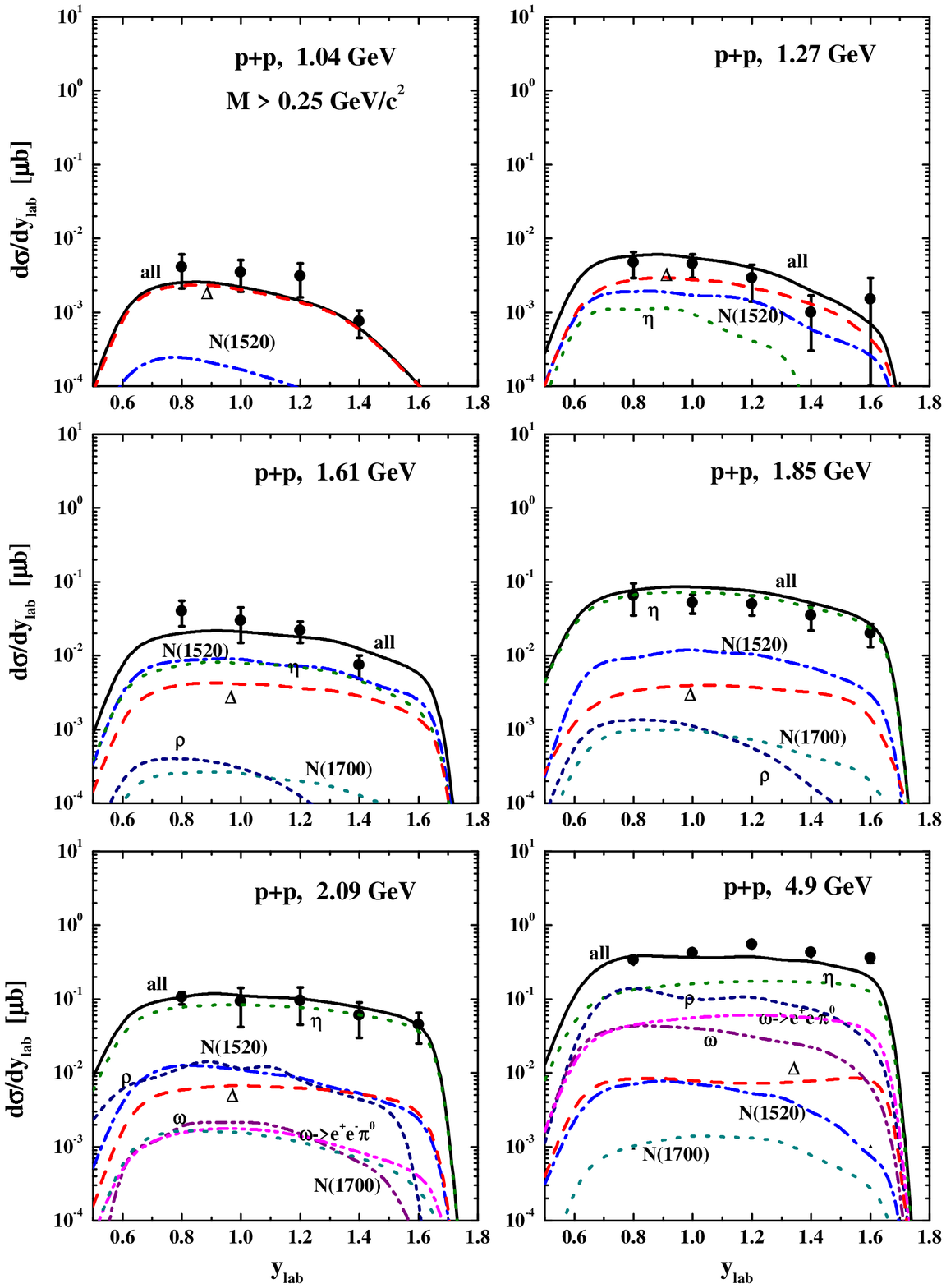,width=15cm}
\caption{The laboratory rapidity spectra of dileptons for $pp$ collisions
at 1.0 -- 4.9~GeV in comparison to the DLS data~\protect\cite{DLSpp}
with a low mass cut of 0.25 GeV/$c^2$.
The assignment of the lines is the same as in Fig. \protect\ref{Fig4pp}.}
\label{Fig6pp}
\end{figure}

In Figs. \ref{Fig5pp} and \ref{Fig6pp} we show the laboratory rapidity
spectra for dileptons from $pp$ collisions at 1.0 -- 4.9 GeV, imposing
as in the data a low mass limit of 0.15 GeV (Fig. \ref{Fig5pp}) and 0.25 GeV
(Fig. \ref{Fig6pp}) in comparison to the DLS data \cite{DLSpp}.
The experimental low mass limits applied
here allow to exclude the contribution of the $\pi^0$ Dalitz decay
and partly suppress (for the 0.25 GeV cut) the contributions from $\eta$
and $\Delta$ Dalitz decays (cf. Fig. \ref{Fig4pp}).

Both at 1.04 and 1.27 GeV the $\Delta$ Dalitz decay is dominant
(Figs. \ref{Fig5pp}, \ref{Fig6pp}).  At 1.61 GeV with a 0.25 GeV
lower mass limit the $N(1520)$ contribution becomes visible (Fig.
\ref{Fig6pp}). However, at 1.85 -- 4.9 GeV basically the $\eta$ Dalitz
decay contributes to the rapidity spectra since this channel has the
largest differential cross section $d\sigma/dM$ (cf. Fig. \ref{Fig4pp}).
Note again, that the $\eta$ cross section from $pp$ collisions is well
known experimentally which provides a valuable control of independent
dilepton data.

\subsection{Excitation function}

\begin{figure}
\epsfig{file=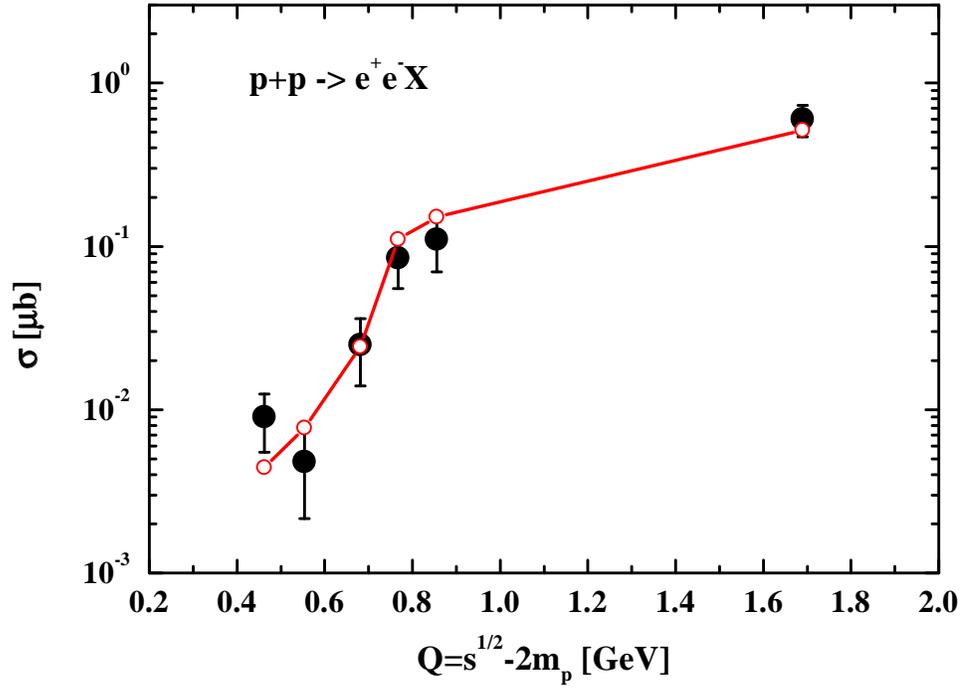,width=15cm}
\caption{The excitation function for dileptons from $pp$ collisions
for masses $M > 0.15$ GeV/$c^2$ in comparison to the DLS
data~\protect\cite{DLSpp}.}
\label{Fig7pp}
\end{figure}

The excitation function, i.e. the integrated dilepton cross section for
masses above 0.15 GeV, is shown in Fig. \ref{Fig7pp} in comparison
to the DLS data \cite{DLSpp} (full circles).
The experimental total cross section at 1.0 GeV
($Q=\sqrt{s}-2m_p$ = 0.46 GeV) is larger than at 1.27 GeV
($Q=0.55$~GeV). We underestimate $\sigma(s)$ at 1.0 GeV, but stay
on the upper level of error bars at 1.27 GeV, which is consistent with
the dilepton mass spectra of Fig. \ref{Fig4pp}.  Thus we get a
monotonous increase of the total cross section with energy because
of the increase of the available phase space and new channels opening
up.  The discrepancy with the DLS data at 1.0 -- 1.27 GeV might be due
to experimental uncertainties, since the excitation function for $pd$
collisions measured by the DLS collaboration indicates also a
monotonous increase \cite{DLSpp}.

\section{Summary}

We have studied  dilepton production from $pp$ collisions at
1.0 -- 4.9 GeV in comparison to the data of the DLS Collaboration
which provide sensible constraints on the different individual
production channels. In addition to the conventional dilepton sources as
$\pi^0$, $\eta$, $\omega$ and $\Delta$ Dalitz decays and direct decays
of vector mesons ($\rho$, $\omega$) we  have included the subthreshold
$\rho$ production via baryonic resonances ($N(1520), N(1700)$).  It has
been shown that the baryonic resonances play an essential role in the
low mass $\rho$ production in pion-nucleon and nucleon-nucleon
collisions below the experimentally seen threshold.  The contribution from
the baryonic resonances, in particular $N(1520)$, can be seen in
the invariant mass dilepton spectra, especially at 1.61 GeV, and in the
rapidity distributions when imposing low mass cuts in order to suppress
the contributions from $\pi^0$, $\eta$ and $\Delta$ Dalitz decays.

Our results for the differential mass spectra including the conventional
sources mentioned above are in general agreement with the calculations
from Ref. \cite{Ernst}. The remaining differences might be attributed to
different parametrizations for the meson production cross sections used
since the existing experimental data so far allow for different fits
within the error bars.

It has been found that the DLS data for $pp$ collisions -- their
invariant mass spectra, laboratory rapidity distributions and total
dilepton production cross section -- can be well described
including all channels mentioned above in an incoherent way. This might
indicate that interference effects between different channels
are hardly visible where integrating over
a wide region in phase space. Our analysis has shown
that the 'input' used in transport calculations for heavy-ion
and proton-nucleus collisions
\cite{Cass95CH,Brat97,CBRW97,BCRW97,BrKo98,CB99PR} is in
agreement with the DLS $pp$ data.

The 'puzzle' that the DLS heavy-ion data cannot be reproduced within such
calculations thus requires future investigations or/and new independent
experimental data as expected in future from the HADES collaboration. We
stress that dilepton data should be taken from $pp$, $pd$, $pA$, $AA$ as well
as $\pi p$, $\pi d$ and $\pi A$ collisions
under the same experimental conditions to allow for a sensitive test of
in-medium properties of the vector mesons.

\acknowledgements
The authors are grateful for valuable discussions with
C. Gale and C. M. Ko.
This work has been supported by GSI, BMBF and DFG.


\newpage

\end{document}